\documentclass[dvips]{article}
\usepackage{icrctc07}

\title{Latest Results from the Air Shower Simulation Programs CORSIKA and CONEX}
\shorttitle{CORSIKA and CONEX results}
\authors{T. Pierog$^{1}$, R. Engel$^{1}$,
   D. Heck$^{1}$, S. Ostapchenko$^{1,2}$ and K. Werner$^{3}$.}
\shortauthors{T. Pierog et al}
\afiliations{$^1$ Forschungszentrum Karlsruhe, Institut f\"ur Kernphysik,
     Karlsruhe, Germany\\ 
$^2$ D.V. Skobeltsyn Institute of Nuclear Physics, Moscow State University, 
     Moscow, Russia\\ 
$^3$ SUBATECH, University of Nantes -- IN2P3/CNRS-- EMN, Nantes, France}
\email{pierog@ik.fzk.de}

\abstract{
Interpretation of EAS measurements strongly depends on detailed air shower simulations. 
The uncertainty in the prediction of shower observables 
for different primary particles and energies is currently dominated by differences 
between hadronic interaction models. 
The new models {\sc qgsjet~II-3} and {\sc epos~1.6}, which reproduce all 
major results of existing accelerator data (including detailed data of RHIC 
experiments for {\sc epos}), have been implemented in the air shower simulation 
programs {\sc corsika} and {\sc conex}. 
We show predictions of these new models and compare them with those 
from older models such as {\sc qgsjet01} or {\sc sibyll}. 
Results for important air shower observables are discussed in detail.
}

\begin{document}
\maketitle

\section{Introduction}

The experimental method of studying ultra-high energy cosmic rays is
an indirect one. Typically, one investigates various characteristics of extensive
air showers (EAS), a huge nuclear-electro-magnetic cascade induced by
a primary particle in the atmosphere, and uses the obtained
information to infer the properties of the original particle, its
energy, type, direction etc.  Hence, the reliability of ultra-high
energy cosmic ray analyses depends on the use of proper theoretical
and phenomenological descriptions of the cascade processes.

The most natural way to predict atmospheric
particle cascading in detail seems to be a direct Monte Carlo (MC) simulation
of EAS development, like it is done, for example, in the {\sc corsika}
program \cite{corsika}. As a very large computation time is required at high energy,
an alternative procedure was developed to describe EAS development numerically,
based on the solution of the corresponding cascade equations.
Combining this with an explicit MC simulation of the most energetic
part of an air shower allows us to obtain accurate results
both for average EAS characteristics and for their fluctuations 
in {\sc conex} program \cite{conex}.

After briefly describing recent changes introduced in {\sc corsika} and {\sc conex}, we will
present the latest results for important air shower observables obtained with 
these models.

\section{Improvements of CORSIKA and CONEX}

Last year {\sc qgsjet~II-3} \cite{qgsjetII} and this year {\sc epos
  1.6} \cite{splitting} have been introduced in both {\sc corsika} and
{\sc conex} as new hadronic interaction models.  These models have
quite different philosophies. The first one is dedicated to cosmic ray
physics and based on the re-summation of enhanced pomeron graphs to
all orders \cite{Ostapchenko:2006vr}. The latter one is designed for
high energy physics and partially relies on a more phenomenological
approach, aiming at a nearly perfect description of accelerator data,
in particular new RHIC measurements.  Some results are presented in
the following (see also \cite{eposmoi}).

\begin{figure}
\begin{center}
\includegraphics [width=0.48\textwidth]{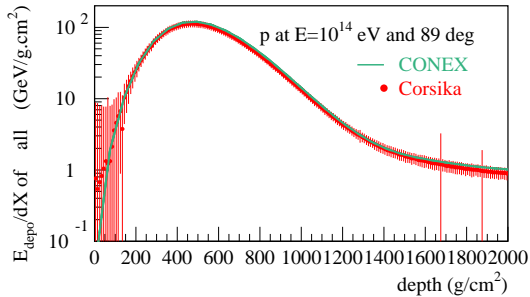}
\end{center}
\caption{Mean longitudinal energy deposit profile in $\rm GeV~g^{-1}~cm^{2}$ as
a function of the slant depth in g cm$^{-2}$ for proton induced $89^{\circ}$ 
inclined showers at $10^{14}~$eV simulated with {\sc corsika} (red dots) and
{\sc conex} (green line) using {\sc qgsjet01}.}\label{fig1}
\end{figure}

Concerning the particle tracking algorithms, the most important
improvement in the last release of {\sc corsika} (6.611) is the
possibility to combine the {\sc slant/upward/curved} options
\cite{heck} in order to simulate air showers with any kind of zenith
angle, including upward going showers (from $0^\circ$ to $180^{\circ}$).  The
calculation of slant depth distances has been improved using the work of
\cite{vladimir} as also employed in {\sc conex}.  In Fig.~\ref{fig1} the
mean longitudinal energy deposit profile is shown as a function of the
slant depth for proton induced showers of $89^{\circ}$ at
$10^{14}$~eV, simulated with {\sc corsika} and {\sc conex} using {\sc
  qgsjet01} \cite{qgsjet01}. Even for this extreme zenith angle, very
good agreement between the two programs is found.

\begin{figure}
\begin{center}
\includegraphics [width=0.48\textwidth]{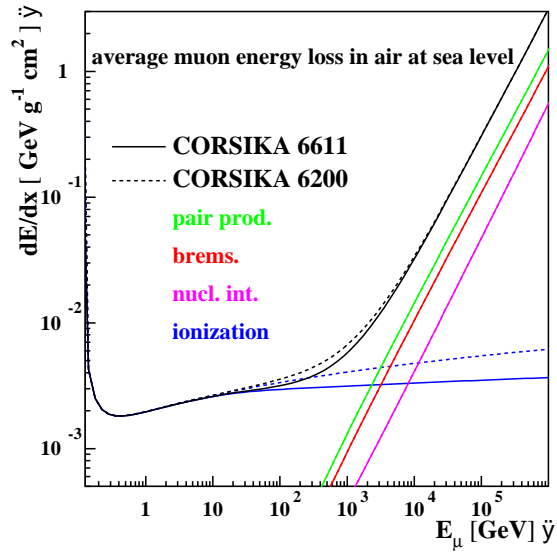}
\end{center}
\caption{Muon energy loss in $\rm GeV~g^{-1}~cm^{2}$ as function of
total muon energy in GeV for {\sc corsika} versions 6.200 (dashed) and 
6.611 (full).}\label{fig2}
\end{figure}

Furthermore, in order to improve muon propagation, the Sternheimer
density correction of the ionization energy loss has been extended to
apply also to muons in both {\sc corsika} and {\sc conex}, based on
work by Kokoulin \& Bogdanov \cite{kokoulin}. The effect of the
density correction can be seen in Fig.~\ref{fig2}.

\begin{figure}
\begin{center}
\includegraphics [width=0.48\textwidth]{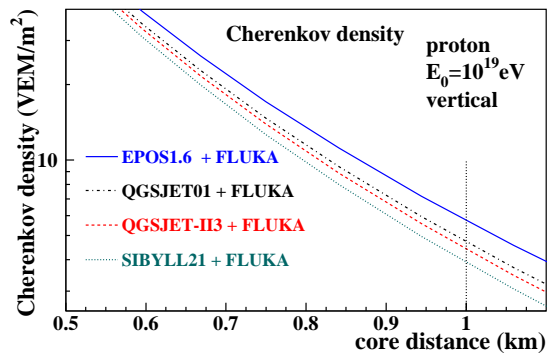}
\end{center}
\caption{Mean lateral distribution function of
Cherenkov density for $10^{19}$~eV vertical proton induced
showers and different high-energy hadronic interaction models, 
{\sc epos~1.6} (full), {\sc qgsjet01} (dashed-dotted), 
{\sc qgsjet~II-3} (dashed) and {\sc sibyll~2.1} (dotted).}
\label{fig3}
\end{figure}

A major technical improvement was achieved in {\sc corsika} by replacing
the old version manager {\sc cmz} by the combination of {\sc autoconf/automake} tools for
the installation and selection of options in {\sc corsika}. 
Compilation has not to be done by the user anymore, rather
Makefiles are generated by {\sc automake}. Options are selected by a 
shell script using {\sc autoconf} and standard C preprocessor commands 
in the {\sc corsika} source code.

Finally, the interfaces to {\sc fluka 2006.3} \cite{fluka} and {\sc herwig~6.51} 
\cite{herwig} have been updated.

\section{Latest results}

In the following air shower simulation results using {\sc epos ~1.6}
and {\sc qgsjet~II-3} are presented and compared to former results
using {\sc qgsjet01} \cite{qgsjet01} or {\sc sibyll~2.1}
\cite{Fletcher:1994bd,engel}.

 In Fig. \ref{fig3}, {\sc corsika}-based estimates for the lateral
 distribution of the Cherenkov signal in Auger tanks \cite{auger} are
 shown. The tank signal has been simulated in a simplified way as only
 the relative differences between the model results are of importance
 here. Due to a much larger muon number at ground in {\sc epos}
 \cite{eposmoi,Pierog:2006qv}, the density at 1\,km shows an excess of
 about 30 to 40\% compared to {\sc qgsjet~II-3} while the latter is
 well in between {\sc qgsjet01} and {\sc sibyll}. Such an excess is of
 crucial importance for the reconstruction of the primary energy and
 composition with the Auger surface detector alone
 \cite{RalphAuger,HolgerKA}. Compared to other models, using {\sc
   epos} would decrease the energy reconstructed from lateral
 densities and could lead to a lighter primary cosmic ray composition.

\begin{figure}
\begin{center}
\includegraphics [width=0.48\textwidth]{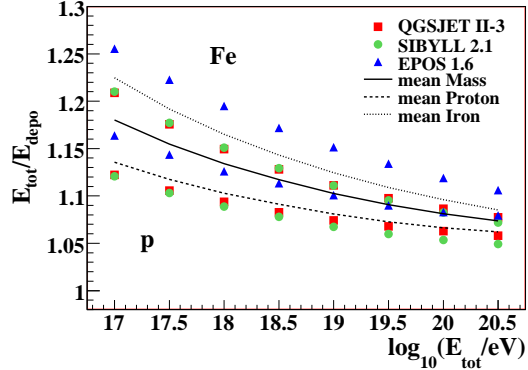}
\end{center}
\caption{\label{factor}Mean factor for the conversion of observed
(calorimetric) energy to total energy for iron (dotted) or proton
(dashed) induced showers. The conversion factor is
shown for {\sc qgsjet~II-3} (circles), {\sc sibyll~2.1} (squares) and
 {\sc epos~1.6} (triangles). The mean conversion factor (full line) is 
calculated by averaging all proton and iron predictions.}\label{fig4}
\end{figure}

 The higher muon number from {\sc epos} is mainly due to a larger
 baryon-antibaryon pair production rate in the individual hadronic
 interactions in showers. By predicting more baryons, more energy is
 kept in the hadronic shower component even at low energy.  As a
 consequence, the calorimetric energy as measured by fluorescence
 light detectors is reduced since more energy is transferred to
 neutrinos and muons.  In Fig. \ref{fig4} the conversion factor from
 the visible calorimetric energy to the real energy is plotted as a
 function of the primary energy of the showers. {\sc qgsjet~II-3} gives results
 very similar to {\sc sibyll}. As expected, {\sc epos} shows a conversion
 factor which is up to 3.5\% higher than other models at low energy.

\begin{figure*}[ht]
\begin{center}
\includegraphics [width=0.75\textwidth]{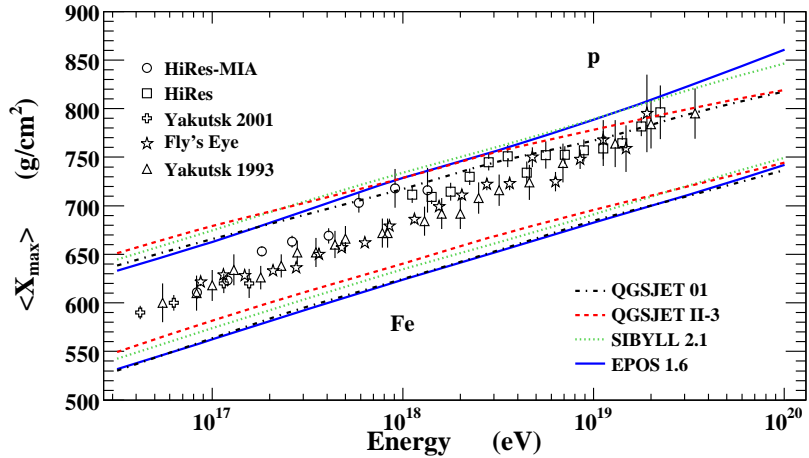}
\end{center}
\caption{Mean X$_{\rm max}$ for proton and iron induced showers as a function of the primary energy. Predictions of different high-energy hadronic interaction models, {\sc qgsjet01} (dashed-dotted), {\sc qgsjet~II-3} (dashed), {\sc sibyll~2.1} (dotted) and {\sc epos ~1.6} (full), are compared to data. Refs. to the data can be found in \cite{Engel:2004ui}}\label{fig5}
\end{figure*}

As shown in Fig.~\ref{fig5}, the mean depth of shower maximum, X$_{\rm
  max}$, for proton and iron induced showers simulated with {\sc
  conex} is nevertheless not very different for {\sc epos}. Up to
$10^{19}~$eV, all models agree within $\rm 20 g~cm^{-2}$. {\sc Epos}
proton induced showers show a slightly higher elongation rate in that
range while {\sc qgsjet~II-3} has a slightly lower one. Above this
energy, both {\sc qgsjet01} and {\sc qgsjet~II-3} elongation rates
decrease due to the very large multiplicity of these models at ultra-high
energy. 
Below $10^{18}~$eV, an analysis of $X_{\rm max}$ data would lead to a composition of
primary cosmic rays that is heavier using {\sc qgsjet~II-3} compared to
{\sc epos}. Above $10^{18}~$eV the situation is reversed.

\section{Conclusions}

New versions of {\sc corsika} and {\sc conex} have been released
recently with two new hadronic interaction models. The models differ
in several important aspects in the approach of reproducing data. In
{\sc qgsjet~II-3}, high parton density effects are treated by
re-summing enhanced pomeron graphs to all orders, but energy
conservation at amplitude level is not implemented. On the other hand,
in {\sc epos}, energy conservation at amplitude level is fully
implemented, but high-density effects are treated by a
phenomenological parametrization. {\sc epos} is particularly
well-tuned to describe available accelerator data including heavy ion
collisions measured at RHIC. The differences of the model predictions
are large: At high energy, proton induced air showers simulated with
{\sc epos} have even more muons at ground than iron induced showers
simulated with {\sc qgsjet~II-3}. Comparison to cosmic ray data, for
example, from the KASCADE detector, are now needed to support or
disfavour the {\sc epos} predictions.

\noindent
{\bf Acknowledgments:}
The {\sc corsika} and {\sc conex} authors would like to thank all users who 
contributed to the development of the programs by helping to detect 
and solve problems. 
We are particularly grateful to R.P. Kokoulin and A.G. Bogdanov for their 
very useful work and comments on muon interactions and energy loss.


\end{document}